\documentclass[12pt]{article}
\usepackage[dvips]{color}                 

\usepackage{graphicx}                     

\usepackage{amssymb}

\usepackage{amsmath}

\usepackage{xspace}                       

\renewcommand{\today}{\ifcase\day\or 1st\or 2nd\or 3rd\or 4th\or 5th\or 6th\or

        7th\or 8th\or 9th\or 10th\or 11th\or 12th\or 13th\or 14th\or 15th\or

        16th\or 17th\or 18th\or 19th\or 20th\or 21st\or 22nd\or 23rd\or 24th\or

        25th\or 26th\or 27th\or 28th\or 29th\or 30th\or

        31st\fi~\ifcase\month\or January\or February\or March\or April\or

        May\or June\or July\or August\or September\or October\or November\or

        December\fi \space \number\year}   

\newcommand{\mytitle}[1]{
                         \begin{center}
                           \LARGE{\textbf{#1}}
                         \end{center}}
\newcommand{\myauthor}[1]{\textbf{#1}}
\newcommand{\myaddress}[1]{\textit{#1}}
\newcommand{\mypreprint}[1]{\begin{flushright} #1 \end{flushright}}

\begin{document}

%

\begin{titlepage}
\mypreprint{
\textbf {TUM-T39-03-03} \\
}

\vspace*{1.0cm}

\mytitle{Quark mass dependence of the nucleon axial-vector coupling constant\footnote{Work supported in part by BMBF and DFG}}


\begin{center}
  \myauthor{Thomas R. Hemmert$^{a}$}, \myauthor{Massimiliano Procura$^{a,b}$}
  and
  \myauthor{Wolfram Weise$^{a,b}$} 

  \vspace*{0.5cm}

  \myaddress{$^a$
    Physik-Department, Theoretische Physik  \\
    Technische Universit{\"a}t M{\"u}nchen, D-85747 Garching, Germany\\
    (Email: themmert@physik.tu-muenchen.de)}\\[2ex]
  \myaddress{$^b$ ECT*, Villa Tambosi, I-38050 Villazzano (Trento), Italy\\
    (Email: weise@ect.it, procura@ect.it)}

  \vspace*{0.2cm}
\end{center}


\begin{abstract}
  We study the quark mass expansion of the axial-vector coupling constant $g_A$ of the nucleon. The aim is to explore the feasibility of chiral effective field theory methods for extrapolation of lattice QCD results---so far determined at relatively large quark masses corresponding to pion masses $m_\pi\gtrsim0.6\,$GeV---down to physical values of $m_\pi$. We compare two versions of non-relativistic chiral effective field theory: One scheme restricted to pion and nucleon degrees of freedom only, and an alternative approach which incorporates explicit $\Delta$(1230) resonance degrees of freedom. It turns out that, in order to approach the physical value of $g_A$ in a leading-one-loop calculation, the inclusion of the explicit $\Delta$(1230) degrees of freedom is crucial. With information on important higher order couplings constrained from analyses of the $\pi N\rightarrow \pi\pi N$ reaction, a chiral extrapolation function $g_A(m_\pi)$ is obtained, which works well from the chiral limit across the physical point into the region of present lattice data. The resulting enhancement of $g_A(m_\pi)$ near the physical pion mass is found to arise from an interplay between long- and short- distance physics. 
\end{abstract}


\noindent


\end{titlepage}

\setcounter{page}{2} 
\newpage

%

\section{Introduction}
Lattice QCD is developing into a powerful tool for studying the structure of nucleons \cite{Schierholz,TW}. In practice, however, these computations are so far limited to relatively large quark masses. The typical ``light'' quark masses that can be managed on the lattice are usually at least ten times larger than the $u$- and $d$- quark masses, $m_{u,d}\simeq 10\,$MeV, determined at renormalization scales around $1\,$GeV. The corresponding masses of a pion on the lattice---defined as the lowest lying $0^-$ state in the simulation---are larger than $0.5\,$GeV, quite far above the physical value $m_\pi\simeq 0.14\,$GeV.

In the limit $m_{u,d}\to 0$, on the other hand, low-energy 2-flavor QCD displays a spontaneously broken chiral $SU(2)\times SU(2)$ symmetry. Pions as massless Goldstone bosons are the relevant degrees of freedom of the (effective) theory. Their coupling to baryons is also subject to the rules imposed by chiral symmetry. It produces the pion cloud of the nucleon, an important low-energy, long-wavelength aspect of nucleon structure, cf.~\cite{TW} and references therein.

For baryon properties the interpolation between lattice results obtained at relatively large pion masses and actual observables determined at the physical $m_\pi$ has become an issue of great interest recently. Chiral effective field theory (ChEFT) can, in principle, provide such extrapolations.
First steps in this direction were made by Leinweber et al. \cite{Leinweber} for the case of the magnetic moments. They used Pad\'e approximants based on the leading dependence on the pion mass as dictated by chiral symmetry. A perturbative analysis of nucleon magnetic moments, using ChEFT with explicit inclusion of the $\Delta$(1230) isobar, has been performed in \cite{HW}. It turned out to be of crucial importance to promote the isovector couplings of the $\Delta$ to leading order in the power counting characteristic of chiral effective theories. 
 Truncated versions of chiral perturbation theory usually ``freeze'' the $\Delta$ and relegate its effects to higher order counterterms, with the consequence that the range of convergence of this truncated theory is often quite limited. The importance of the $\Delta$ as an explicit degree of freedom in various aspects of nucleon substructure has long been known. It has played a prominent part in chiral models of the nucleon \cite{TW,tony} and earlier studies based on current algebra \cite{pagels}.

In the present paper\footnote{Some aspects of this work have already been reported in \cite{conferences}.} we investigate the quark mass dependence of $g_A$, the axial-vector coupling constant of the nucleon. This is a key quantity for our understanding of the nucleon's chiral structure \cite{TW}. Furthermore, lattice QCD determinations of $g_A$ are progressing to the point where such an investigation is in demand for interpolating between lattice data and the actual observable \cite{Schierholz,data}. 

Of course plenty of ChEFT calculations pertaining to this quantity exist in the literature \cite{BKM}, and the special importance of intermediate spin 3/2 resonance contributions for the axial properties of baryons has been recognized a long time ago. 
Recall, for example, the Adler-Weisberger sum rule \cite{AW}:
\begin{eqnarray}
g_A^2&=&1+\frac{2f_\pi^2}{\pi}\int_{m_\pi}^{\infty}\frac{d\omega}{\sqrt{\omega^2-m_\pi^2}}[\sigma_{\pi^+ p}(\omega)-\sigma_{\pi^- p}(\omega)]+{\cal O}\left(\frac{m_\pi^2}{M_N^2}\right)\;.\label{AW}
\end{eqnarray}
It relates the surplus of the axial-vector coupling constant beyond its ``trivial'' value $g_A\,=\,1$ (for a point-like, structureless nucleon) to the excess of the $\pi^+ p$ cross section over the $\pi^- p$ cross section, a feature that is dominated by $\Delta\,(1230)$ resonance excitation. 

In ChEFT $g_A$ is calculated from the response of a baryon to the presence of an external background field with axial quantum number.
In contrast to many earlier ChEFT calculations---{\it e.g.} the pioneering ones of refs.\cite{BSW,JM}---we are {\em not only} interested in the non-analytic quark mass behaviour generated by the Goldstone boson cloud around the baryon. In order to analyze the quark mass dependence of a baryon observable in a quantitative way we follow the same philosophy as spelled out in ref.\cite{HW}. We specify a power counting scheme plus a certain order in that scheme and then systematically evaluate all contributions---short- and long-distance---to that order. It turns out that the role of short-distance physics parameterized as higher order operators in chiral effective field theory is crucial for an understanding of the chiral extrapolation function $g_A(m_\pi)$. We utilize the so-called SU(2) ``small scale expansion'' of ref.\cite{review} to leading-one-loop order (${\cal O}(\epsilon^3)$), which includes explicit nucleon and $\Delta$(1230) degrees of freedom---some details regarding the formalism are summarized in section \ref{2}. In fact, all the long- and short-distance contributions considered in our analysis have already been found (as a by-product) in ref.\cite{BFHM}, where among other topics the (small) momentum dependence of the axial form factors of the nucleon was studied. Here we reconsider these results and focus on the quark mass dependence of $g_A$. Unlike the situation in the vector current sector, where a modified power counting had to be developed \cite{HW} so that important effects of the $\Delta$ could be captured already in leading-one-loop order, this turns out not to be necessary for axial current effects related to the $\Delta$(1230). For the {\em axial structure of baryons}---which is the topic of the present analysis---the standard counting of ChEFT, as employed in refs. \cite{review,BFHM}, is sufficient to obtain $\Delta$(1230) induced quark mass dependence already at the leading-one-loop level. Throughout this work we can therefore apply standard (``naive'') counting rules.

Historically, attempts to obtain a chiral extrapolation function for $g_A$ based on the known leading-non-analytic (LNA) quark mass dependence in combination with a phenomenological (quark mass dependent) regularization procedure did not yield satisfactory results, displaying axial couplings less than unity at the physical point \cite{Detmold1}. In our opinion one should not have been surprised about such a failure, as the LNA quark mass term presumably is only dominant for quark masses near the extreme chiral limit of the theory. Such a feature was indeed observed in the analysis of the anomalous magnetic moment of the nucleon \cite{HW}. Recently, Detmold et al. \cite{Detmold2} in their analysis of moments of polarized DIS structure functions, found an improved extrapolation formula for $g_A(m_\pi)$ utilizing a chiral quark model which also allows for contributions from intermediate $\Delta$(1230) states \cite{tony}. However, the resulting extrapolation function---which has most of the $\Delta$(1230) related couplings fixed from SU(6) symmetry---still does not provide for an enhancement of $g_A(m_\pi)$ near the physical point, which would at the same time connect with the lattice results. In this work we go beyond the existing analyses by performing a systematic perturbative calculation of {\em all} short- and long-distance contributions which are allowed to leading-one-loop order in chiral power counting. 

This paper is organized as follows. In the next section we give a brief summary of the ChEFT formalism utilized here. In section \ref{3} we present the leading-one-loop analytic result for $g_A$ both in a scheme without and in one with explicit $\Delta$(1230) degrees of freedom. In section \ref{4} we compare these analytic results with recent data from the QCDSF collaboration \cite{gadata}, employing three different fit procedures. Finally, in section \ref{5} we check the stability of the obtained results by including the physically known information about $g_A$ as an additional constraint in the fit. The resulting chiral extrapolation function $g_A(m_\pi)$ provides a sensible description of the quark mass dependence from the chiral limit over the physical point out to the present lattice data. We conclude with an outlook on further investigations required for a better understanding of $g_A$.

\section{Chiral Effective Field Theory Input}\label{2}

Our analysis of $g_A$ is based on chiral effective field theory with two light flavors $(u,d)$. We work to leading-one-loop order. All other quark degrees of freedom are integrated out, leaving their marks only in slightly shifted values of couplings sensitive to short distance dynamics. This effective field theory of low-energy QCD acts with pions as Goldstone bosons of the spontaneously broken chiral $SU(2) \times SU(2)$ symmetry of QCD. In addition the chiral symmetry of the QCD Lagrangian with two light flavours is explicitly broken by the small $u-$ and $d-$ quark masses which are treated as a perturbation. They shift the massless Goldstone bosons to the physical pions with mass $m_\pi$ and lead to a string of quark mass dependent operators which turn out to be of crucial importance for the understanding of chiral extrapolation functions.  At present we ignore effects of isospin breaking and work with degenerate masses, $\hat{m}=(m_u+m_d)/2$, for the up and down quarks. 

We utilize two versions of ChEFT for our analysis of $g_A$. The first scheme is SU(2) heavy-baryon chiral perturbation theory (HBChPT), which involves only the Goldstone boson modes (pions) and spin 1/2 matter fields (nucleons), taken up to order $p^3$. The second scheme is the SU(2) small scale expansion (SSE) which in addition also involves explicit spin 3/2 matter fields (four $\Delta$(1230) states) and their interactions with Goldstone bosons, taken up to order $\epsilon^3$ in the ``small scale'' $\epsilon$ which now includes also the non-zero $\Delta$-$N$ mass difference \cite{letter}.

For completeness---and for the proper definition of our couplings---we give the effective Lagrangian required for this leading-one-loop analysis of $g_A$:
\begin{eqnarray}
{\cal L}={\cal L}^{(1)}_{N}+{\cal L}^{(2)}_N+{\cal L}^{(3)}_{N}+{\cal L}^{(1)}_{N\Delta}+{\cal L}^{(1)}_{\Delta}+{\cal L}^{(2)}_{\pi\pi}\;.
\end{eqnarray}
The well-known leading order pion Lagrangian ${\cal L}_{\pi\pi}^{(2)}$
can be found in \cite{GL}. The leading order $\pi N$, $\pi N \Delta$ and $\pi \Delta$ Lagrangians
\begin{eqnarray}
{\cal L}^{(1)}_{N}&=&\bar{N}_v\left[i\,v\cdot D+g_A^0\,S\cdot u\right]N_v
\; ,\nonumber\\
{\cal L}^{(1)}_{N\Delta}&=&\bar{T}^\mu_i\,c_A\,w_{\mu}^i\;N_v\,+\,h.c.\; ,
\nonumber \\
{\cal L}^{(1)}_{\Delta}&=&\,\bar{T}_i^\mu\left[\xi^{ij}_{I=3/2}\,\Delta_0 -iv\cdot D^{ij}-g_1\,S\cdot u\,\delta^{ij}\right]g_{\mu\nu}\,T_j^\nu\; 
\label{LO}
\end{eqnarray}
are specified in detail in \cite{review}. Here we only note that $N_v$ ($T^\mu_i$) represents the non-relativistic spin-1/2 nucleon (spin 3/2 delta) field, with $\xi_{I=3/2}^{ij}$ denoting the isospin 3/2 projection operator. The chiral tensors $u_\mu,\,w_\mu^i$ encode couplings to (external) axial sources with $NN$-, $N\Delta$- and $\Delta\Delta$- axial coupling constants $g_A^0,\,c_A$ and $g_1$, defined in the chiral limit. Furthermore, $D_\mu,\;D_\mu^{ij}$ denote the chiral covariant derivatives of the nucleon, respectively the Delta. $\Delta_0$ corresponds to the $N\Delta$ mass splitting in the chiral limit. The two four-vectors $v_\mu,\,S_\mu$---related to kinematics and spin---are discussed in the Appendix. It can be clearly seen from Eq.(\ref{LO}) that all couplings considered as leading order follow the standard counting of ChEFT.

While the leading order Lagrangians given above are well-known in the literature we also want to discuss the less-known higher order couplings required for a complete leading-one-loop calculation of $g_A$ both in HBChPT and in SSE. We utilize
\begin{eqnarray} 
{\cal L}^{(2)}_N&=&\bar{N}_v\left[-\frac{i\,g_A^0}{2 M_0}\left\{S\cdot D,v\cdot u\right\}_+ +\ldots\right]N_v \;,\nonumber\\
{\cal L}^{(3)}_{N}&=&\bar{N}_v\left[\frac{g_A^0}{4M_0^2}\,v\cdot D\,S\cdot u\,v\cdot D-\frac{g_A^0}{4M_0^2}\,\left(\left\{S\cdot D,v\cdot u\right\}v\cdot D+h.c.\right)\right.\nonumber\\
& &
-\frac{g_A^0}{8M_0^2}\,\left(S\cdot u\,D^2+h.c.\right)+\frac{g_A^0}{4M_0^2}\,\left(S\cdot D\,u\cdot D+h.c.\right)
+B_9\,S\cdot u\;{\rm Tr}(\chi_{+})\nonumber\\
& &
+B_{20}\left[{\rm Tr}(\chi_+)\;iv\cdot D+h.c.\right]+\Delta_0^2\, B_{30}\,iv\cdot D+\Delta^2_0\, B_{31}\,S\cdot u+\ldots \Bigr] N_v\;.\label{counterterms}
\end{eqnarray}
Here $M_0$ denotes the nucleon mass in the chiral limit, whereas the chiral tensor $\chi_+$ encodes quark-mass dependent short distance physics. For the higher order couplings $B_i$ we follow the nomenclature of ref.\cite{BFHM} where a complete set of counter terms is listed, appropriate for renormalization of both leading one-loop HBChPT and SSE calculations. We note that $B_{30}$ and $B_{31}$ are identically zero in HBChPT, but are required for leading-one-loop calculations in SSE. In general all $B_i$ have a finite regularization scale $\lambda$ dependent part $B_i^{r}(\lambda)$ as well as an infinite part: 
\begin{eqnarray}
B_i\equiv B_i^{r}(\lambda )+\frac{\beta_i}{(4\pi f_\pi)^2}\; 16 \pi^2 \,L\;.
\end{eqnarray}
In the infinite part $\beta_i$ denotes the HBChPT (respectively SSE) beta-function associated with this counterterm, whereas the infinities are encoded in the quantity $L$, discussed in the Appendix. $f_\pi$ is the pion-decay constant.

\section{Analytic Results}\label{3}

\begin{figure}[t]
  \begin{center}
    \includegraphics*[width=0.7\textwidth]{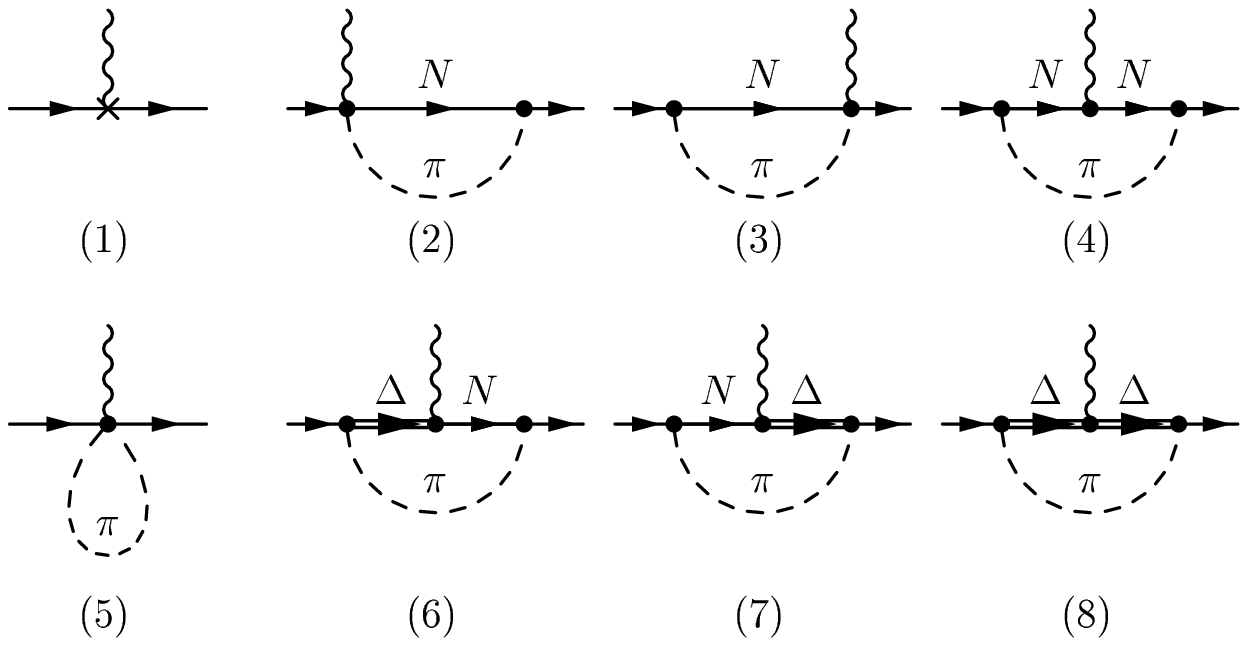}
    \caption{Diagrams contributing to the nucleon axial-vector coupling constant $g_A$ at leading-one-loop order. The wiggly line denotes an external (isovector) axial-vector background field, interacting with a nucleon (solid lines). }
    \label{Feyndiags}
  \end{center}
\end{figure}

We now proceed to a discussion of the results for the axial coupling of the nucleon expressed as a function of the pion mass $g_A(m_\pi)$, first for the HBChPT and then for the SSE case. The leading one-loop HBChPT result for the quark mass dependence of $g_A$ is known for a long time \cite{BKM}. Evaluating diagrams 1-5 of Fig.\ref{Feyndiags} and projecting out the axial coupling constant gives
\begin{eqnarray}
g_A(m_\pi^2)&=&g^0_A\,Z_N+4m_\pi^2\;B_9^r(\lambda )+\frac{m_\pi^2}{32\pi^2 f_\pi^2}\left[((g^0_A)^3-4g^0_A)\ln{\frac{m_\pi}{\lambda }}+(g^0_A)^3\right]+{\cal O}(p^4)\;,
\end{eqnarray}
where $g^0_A$ again denotes the axial-vector coupling constant in the chiral limit. The Z-factor of the nucleon to leading-one-loop order reads
\begin{eqnarray}
Z_N(m_\pi^2)&=&1-\frac{3(g^0_A)^2 m_\pi^2}{32\pi^2 f_\pi^2}\left(1+3\ln{\frac{m_\pi}{\lambda}}\right)-8 m_\pi^2 B_{20}^r(\lambda)+{\cal O}(p^4)\;.
\end{eqnarray}
Following ref.\cite{Ecker}, we note that this Z-factor is finite since we have not transformed away the counter term $B_{20}$. It is well-known, however, that the finite part of a low-energy counter term such as $B_{20}$ cannot be observed independently (of $B_9^r(\lambda)$ in our case). To order $p^3$ in HBChPT, $g_A(m_\pi^2)$ therefore depends on only two unknown parameters, $g_A^0$ and $C(\lambda)\equiv B_9^r(\lambda)-2g^0_AB_{20}^r(\lambda)$, for each choice of the regularization scale $\lambda$:
\begin{eqnarray}
g_A^{HB}(m_\pi^2)&=&g^0_A-\frac{(g^0_A)^3}{16\pi^2f_\pi^2}\,m_\pi^2+4\left\{C^{HB}(\lambda)+\gamma^{HB}\ln{\frac{m_\pi}{\lambda }}\right\}m_\pi^2+{\cal O}(p^4)\;,\label{gAA}
\end{eqnarray}
with
\begin{eqnarray}
\gamma^{HB}&=&-\frac{1}{16\pi^2 f_\pi^2}\left[(g^0_A)^3+\frac{1}{2}\,g^0_A\right]\;.
\end{eqnarray}
We expect Eq.(\ref{gAA}) to hold for sufficiently small quark masses, close to the chiral limit. All observable short distance dynamics is collected in $C(\lambda)$.
The procedure outlined here is well-defined in the sense that the scale-dependent logarithm in Eq.(\ref{gAA}) cooperates with the scale-dependent coefficient $C(\lambda)$ in just such a way that the overall sum is scale-independent.

Next we turn to the result of the SSE approach which includes explicit $\Delta\,(1230)$ degrees of freedom. The bare result depending on all four counterterms of Eq.(\ref{counterterms}) has been given in ref.\cite{BFHM}. It originates from diagrams 1-8 of Fig.\ref{Feyndiags}. We note that diagrams 2 and 3 in principle also exist when the intermediate nucleon is replaced by a spin 3/2 baryon. However, the leading order $N\Delta$ transition Lagrangian of Eq.(\ref{LO}) does not include operators connecting $N\Delta$ via an even number of axial fields. The corresponding diagrams are therefore of higher order than the ones considered here at the leading-one-loop level. Extending the work of ref.\cite{BFHM}  we also impose the constraint of decoupling which implies that, in the chiral limit, effects of the $\Delta$(1230) can be absorbed in contact interactions. This makes sure that the addition of explicit Delta degrees of freedom maintains the correct $m_\pi \to 0$ limit of $g_A$.
\begin{eqnarray}
B_{30}^r(\lambda)&=&\frac{1}{\pi^2 f_\pi^2}\;c_A^2\ln{\frac{2\Delta_0}{\lambda}}\nonumber\\
B_{31}^r(\lambda)&=&\frac{c_A^2}{\pi^2 f_\pi^2}\;\left(\frac{40}{243}g_1-\frac{16}{81}g^0_A+\frac{25}{81}g_1\ln{\frac{2\Delta_0}{\lambda }}+\frac{8}{27}g^0_A\ln{\frac{2\Delta_0}{\lambda}}\right) \label{decoupling}
\end{eqnarray}
and the leading-one-loop SSE Z-factor given in the Appendix we find the chiral limit behaviour
\begin{eqnarray}
g_A^\chi(m_\pi^2)&\approx&g^0_A+4\,\gamma^{HB}\,m_\pi^2\ln
                          \frac{m_\pi}{\lambda}+m_\pi^2\left\{
          -\frac{(g^0_A)^3}{16\pi^2f_\pi^2}+4\,C^{SSE}(\lambda)
          \right.\nonumber\\
 & &\left.+\frac{c_A^2}{\pi^2 f_\pi^2}\left[
          \left(\frac{25 g_1}{162}
          -\frac{g_A^0}{18}\right)\ln\frac{2\Delta_0}{\lambda}
          +\frac{115}{486}\,g_1
          -\frac{35}{54}\,g_A^0\right]\right\}+{\cal O}(m_\pi^3)\;.
          \label{gachiral}
\end{eqnarray}
We emphasize the point that, while the result (\ref{gachiral}) is unique, the separation into $B_{30}^r$ and $B_{31}^r$ can be done in several possible ways.
Comparing with Eq.(\ref{gAA}) we note that with our parameter choice of Eq.(\ref{decoupling}) we have restored $g_A^0$ as the chiral limit value of $g_A(m_\pi^2)$. Furthermore, we note that the leading quark mass dependence starting proportional to $m_\pi^2\ln m_\pi$ is not modified by the addition of explicit $\Delta$(1230) degrees of freedom, again consistent with chiral symmetry. 

Having taken care of the decoupling constraints we now present the complete leading-one-loop result for $g_A$ in the Small Scale Expansion:
\begin{eqnarray}
g_A^{SSE}(m_\pi^2)&=&g^0_A-\frac{(g^0_A)^3m_\pi^2}{16\pi^2f_\pi^2}
                     +4\left\{C^{SSE}(\lambda)
                     +\frac{c_A^2}{4\pi^2 f_\pi^2}\left[\frac{155}{972}\, g_1
                     -\frac{17}{36}\, g^0_A\right]+\gamma^{SSE}
                     \ln{\frac{m_\pi}{\lambda }}\right\}m_\pi^2\nonumber\\
                  & &+\frac{4c_A^2 
                     g^0_A}{27 \pi f_\pi^2 \Delta_0}m_\pi^3
                     +\frac{8}{27\pi^2 f_\pi^2}\;c_A^2 g^0_A m_\pi^2
                     \sqrt{1-\frac{m_\pi^2}{\Delta_0^2}}\,\ln{R}\nonumber\\
                  & &+\frac{c_A^2\Delta_0^2}{81\pi^2 f_\pi^2}\left(25g_1-
                     57g_A^0\right)\left\{\ln\left[\frac{2\Delta_0}{m_\pi}
                     \right]-\sqrt{1-\frac{m_\pi^2}{\Delta_0^2}}\ln R\right\}
                     +{\cal O}(\epsilon^4)\;,\label{gasse}
\end{eqnarray}
with
\begin{eqnarray}
\gamma^{SSE}&=&\frac{1}{16\pi^2 f_\pi^2}\left[\frac{50}{81}\,c_A^2 g_1-\frac{1}{2}\,g_A^0-\frac{2}{9}\,c_A^2g_A^0-(g_A^0)^3\right]\;,\nonumber\\
R&=&\frac{\Delta_0}{m_\pi }+\sqrt{\frac{\Delta_0^2}{m_\pi^2}-1}\,.\label{R}
\end{eqnarray}
We note that the first line in Eq.(\ref{gasse}) has the same structure as the heavy baryon result of Eq.(\ref{gAA}). New structures appearing in the leading-one-loop SSE result are the terms proportional to $m_\pi^3$ as well as the logarithms depending explicitly on the N-$\Delta$ mass splitting value. For the chiral extrapolation of $g_A$ we will utilize the full analytic form as given in Eq.(\ref{gasse}). The much simpler chiral limit form Eq.(\ref{gachiral}) will only be used to constrain coupling constants.

\section{Chiral Extrapolation to Lattice Data}\label{4}

\subsection{General Remarks}

We now turn to a numerical evaluation of the two analytic extrapolation formulae for $g_A$ considered here. In principle all couplings and masses---aside from $m_\pi$---on the left hand side of Eqs.(\ref{gAA},\ref{gasse}) are understood to denote their values in the chiral limit. However, for many of them the chiral limit values are only poorly known. For the pion decay constant we use its physical value $f_\pi$, because the difference between this and $f_\pi^0$ is known to be only a few percent. For $c_A$ and $\Delta_0$---following the reasoning laid out in ref.\cite{HW}---we also use the empirical values, specified in table \ref{tab1}. 

\begin{table}[t]
\caption{Values of the parameters which are taken at their physical value\label{tab1}}
 \begin{center}
  \begin{tabular}{|c|c|}\hline 
  Parameter & Physical Value  \\ \hline \hline
  $f_\pi^0\rightarrow f_{\pi}$ & 0.0924 GeV  \\ \hline
  $c_A^0\rightarrow c_A$ & 1.125  \\ \hline
  $\Delta_0\rightarrow Re[\Delta]$ & 0.2711 GeV \\ \hline  
  \end{tabular}
 \end{center}
\end{table}

Choosing $\lambda=1\,{\rm GeV}$ without loss of generality in leading-one-loop SSE (HBChPT), we are then left with three (two) poorly known couplings: $g^0_A$, $C^{SSE}(1\,{\rm GeV}),\,g_1$, or $g^0_A,\,C^{HB}(1\,{\rm GeV})$, respectively. To determine these short-distance physics parameters of interest, we utilize the (quenched) $g_A$ simulation data of the QCDSF collaboration \cite{gadata} in the mass region $0.3\, {\rm GeV}^2<m_\pi^2<0.6\, {\rm GeV}^2$, containing 5 data points. Hereby we are working under the {\em assumption}---following again the reasoning of ref.\cite{HW}---that for lattice pion masses larger than 600 MeV the effects of ``quenching'' can be neglected. This intuitive expectation has recently been put on firm ground by the analysis of ref.\cite{data}. It was concluded that for present state-of-art simulation data for moments of nucleon structure functions (such as $g_A$) indeed no significant differences between quenched and fully dynamical QCD simulations could be found for $m_\pi>500$ MeV. However, the difference\footnote{In quenched QCD the axial coupling of the nucleon develops a chiral singularity $g_A(m_\pi\rightarrow 0)\sim\log m_\pi$ in the chiral limit \cite{quenched}, in contrast to the finite value $g_A(m_\pi\rightarrow 0)\rightarrow g_A^0$ in full QCD.} between ``quenched'' and fully-dynamical QCD simulations should become visible at lower pion masses. 

While we are confident that (un)quenching effects are small in the mass range considered here possible corrections arising from the finite simulation volume could turn out to be important for the exact position of the lattice data. The five data points we are utilizing here actually result from three different lattice spacings which, however, roughly correspond to the same physical volume\footnote{While this volume at first glance appears rather small to study the physics of an extended object like a nucleon, we note that the nucleon radii are considerably smaller at these large quark masses than at the physical point. See ref.\cite{future} for an analysis of the quark mass dependence of the electromagnetic radii.} of (L=1.45\ldots 1.60 fm)$^3$ \cite{gadata}. Given this data situation no conclusions on finite volume dependence can be drawn from the QCDSF data. In ref.\cite{RBC} for a simulation utilizing domain wall fermions it has been reported that $g_A$ is a quantity which can be sensitive to such corrections. We note that the data analyzed here are consistent with one another (c.f. Fig.\ref{su6fit}) and are basically flat in their $m_\pi$ dependence. The data shown in ref.\cite{RBC} are also rather flat for the largest volume considered there, albeit the overall location of the plateau seems to lie slightly higher than the QCDSF data discussed here. At present no systematic study of these effects exists and we therefore assume that possible shifts in the data due to the finite simulation volume fall within the size of the error bars of the data. Once a larger amount of data for $g_A(m_\pi)$ are obtained at different lattice sizes, we will take up this issue again to recheck our analysis.  

For the purpose of our numerical analysis the $g_A$ simulation data thus only provide the constraint of specifying the location of the typical ``data plateau'' ({\it e.g.} see Fig.\ref{su6fit}), which is quite analogous to the situation in the case of the anomalous magnetic moments \cite{HW}. Once more, we emphasize the empirical observation that any sensible chiral extrapolation formula must have enough structure to reproduce such a plateau, at least over a certain range in $m_\pi^2$. In essence, this implies a strong constraint: different terms in the quark mass expansion of $g_A$ must co-operate in such a way that the pion cloud effects are basically counterbalanced by short-distance physics once the pion mass exceeds $0.6\,$GeV. Given that the available lattice data are still restricted to relatively large quark/pion masses, can one nevertheless utilize this information to determine the chiral extrapolation function? In order to answer this question we perform three different fit procedures of varying sophistication.

\subsection{Free Fits}

\begin{figure}[t]
  \begin{center}
    \includegraphics*[width=0.7\textwidth]{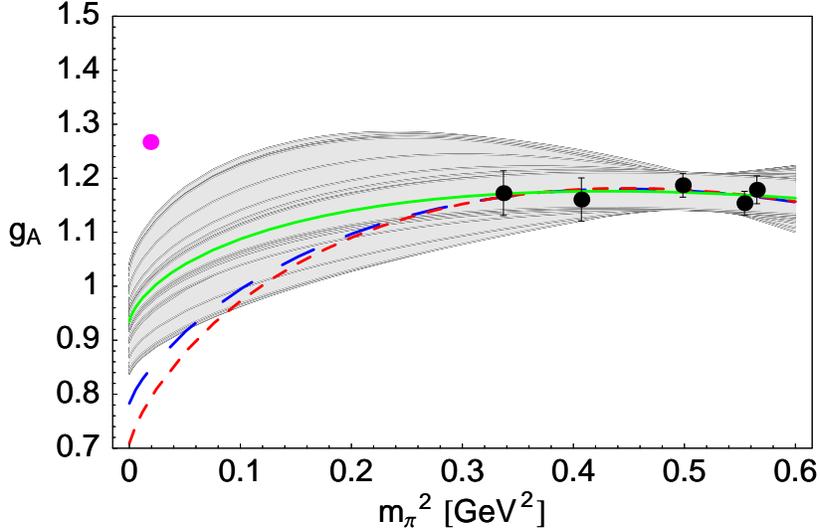}
    \caption{The long-dashed curve represents the free fit to the QCDSF lattice data below $m_\pi=750$ MeV, utilizing the leading-one-loop SSE (Fit Ib) results of Eq.(\ref{gasse}). The analogous leading-one-loop HBChPT result (fit Ia) originating from Eq.(\ref{gAA}) is shown as the short-dashed curve. The solid line represents fit II for the SSE extrapolation with the additional SU(6) quark model constraint $g_1=9/5\,g_A^0$. The indicated error band results from the 95\% confidence ellipse shown in Fig.\ref{su6confidence}, while the solid dot indicates the physical value of $g_A$.}\label{su6fit}
  \end{center}
\end{figure}

In the first round of fits we want to explore whether the available lattice data are sufficient to constrain the unknown parameters $g^0_A$, $C^{SSE}(\lambda),\,g_1$ for SSE, or respectively, $g^0_A,\,C^{HB}(\lambda)$ for HBChPT. The numerical values obtained are given in table \ref{tab2} with the labels fit Ia, fit Ib. The results are shown in Fig.\ref{su6fit} via the short-dashed (HBChPT) and long-dashed (SSE) curves. While the $\chi^2$ per degree of freedom (c.f. table \ref{tab2}) for either curve is small, clearly both extrapolation functions have to be considered {\em unphysical}, as indicated by the small values of $g_A^0$ violating the Adler-Weisberger sum rule as well as the physical constraint $g_A^{phys.}=1.2670(30)$ obtained from neutron beta decay \cite{betadecay}. In addition, the SSE curve has a value of the axial $\Delta\Delta$ coupling constant $g_1\approx 0$, whereas we expect $g_1\geq g_A^0$ from SU(6) symmetry considerations. We note that the resulting (unphysical) HBChPT curve is quite similar to one found in ref.\cite{Detmold1}, where the chiral logarithm with $\gamma^{HB}$ of Eq.(\ref{gAA}) plus a phenomenological (quark mass dependent) cutoff procedure parameterizing short-distance physics was employed. We note that both the HBChPT as well as the SSE ``free fits'' prefer a small positive value for the short distance physics parameterized via $C(\lambda)$---at a regularization scale of $\lambda=1$ GeV. The reason for this unsatisfactory situation is that the present lattice data are not yet sufficiently accurate to obtain a realistic value for $C$, as we will see in the next two sections: additional physics constraints need to be invoked in order to limit the range of $C$.

\begin{table}[t]
\caption{Fit results discussed in the text\label{tab2}}
 \begin{center}
  \begin{tabular}{|c|c|c|c|c|c|}\hline 
  Fit & $g^0_A$ & $C(\lambda=1{\rm GeV})\;\left[{\rm GeV}^{-2}\right]$ & $g_1$ & d.o.f. &$\chi^2$/d.o.f.  \\ \hline \hline
  Ia & $0.71\pm 0.04$ & $+0.12\pm 0.03$ & - & 3 &0.37 \\\hline 
  Ib & $0.78\pm 0.04$ & $+1.06\pm 0.08$ & $0.0\pm 0.1$ &2 & 0.55 \\\hline
  II & $0.94\pm 0.04$ & $-0.25\pm0.04$ &  9/5 $g_A^0$ & 3&0.39 \\\hline
  III & $1.12\ldots 1.26$ & $-2.2\ldots -4.6$  & $4.3\ldots 6.6$   & 3&$0.48\ldots 0.60$ \\\hline
  IVa & $1.36\pm0.07$ & $-5.3\pm1.2$ &  -  & - & -\\\hline
  IVb & $1.21\pm 0.01$ & $-3.4\pm 0.4$ & $5.6\pm 0.5$ & 3 &0.54  \\\hline
  IVc & 1.21 & $-5.3\pm1.2$  & -   & - & -\\\hline
  \end{tabular}
 \end{center}
\end{table}

\subsection{SU(6) Constrained Fit}

\begin{figure}[t]
  \begin{center}
    \includegraphics*[width=0.7\textwidth]{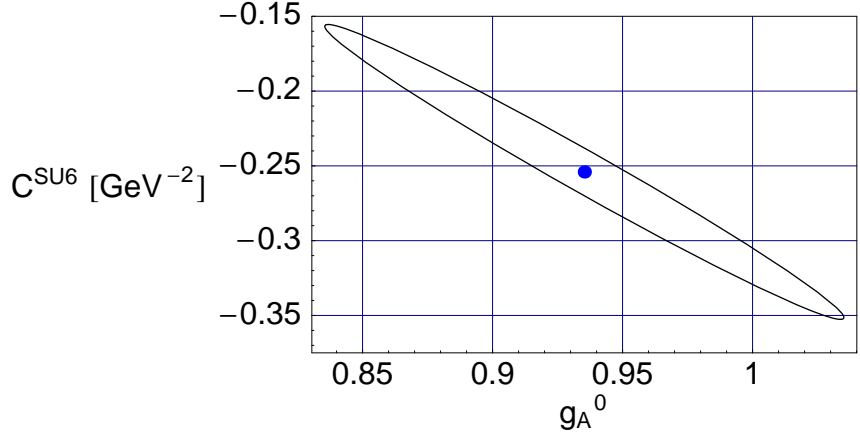}
    \caption{Confidence ellipse for the SU(6)-constrained leading-one-loop SSE fit II shown in Fig.\ref{su6fit}. The two parameters shown denote the nucleon coupling constant $g_A^0$ in the chiral limit versus the parameter $C(\lambda)$ discussed in the text, evaluated at the regularization scale $\lambda=1$ GeV.}\label{su6confidence}
  \end{center}
\end{figure}

One possible conclusion from the mismatch between the chiral extrapolation and present lattice data could be that the extrapolation formulae (\ref{gAA}) and (\ref{gasse}) are just too simplistic to cover such a large region in $m_\pi^2$ (for example, this was the case with the corresponding extrapolation functions of schemes A or B of the anomalous magnetic moments in ref.\cite{HW}). In the following we argue that this is indeed true for the leading-one-loop HBChPT result of Eq.(\ref{gAA}), whereas the corresponding SSE extrapolation function Eq.(\ref{gasse}), with the $\Delta$(1230) added as an explicit degree of freedom, does contain enough quark-mass dependent structures to reproduce the plateau-like behaviour at pion masses above 600 MeV without missing the physical value of $g_A$. In other words, we suspect that the lattice data are just not accurate enough to properly constrain all three parameters $g^0_A$, $C^{SSE}(1\,{\rm GeV}),\,g_1$---which does not come as complete surprise given the low statistics of our data sample. In order to test this hypothesis we are trying to further constrain these couplings via available information from hadron phenomenology. An obvious idea is to fix the axial $\Delta \Delta$ coupling $g_1=9/5\, g_A^0$ from SU(6) symmetry, leaving only $g_A^0$ and $C$ as free parameters to be determined from the fit utilizing Eq.(\ref{gasse}). The numerical values resulting from this fit are shown in table \ref{tab2} with the label fit II. The associated chiral extrapolation curve is shown in Fig.\ref{su6fit}. Its $\chi^2$/d.o.f. is comparable with the values for the free fits Ia and Ib. We conclude that---from a physical point of view---the SU(6) constrained fit does slightly better than the free fits when one considers the ``higher'' value for $g_A^0$. Nevertheless, the axial coupling is still smaller than 1 in the chiral limit, which is hard to reconcile with the Adler-Weisberger sum rule Eq.(\ref{AW}). Surprisingly, even when an error analysis at the 95\% confidence level is performed (see Fig.\ref{su6confidence}), the resulting band shown in Fig. \ref{su6fit} does not include the physical point, despite the rather large error bars of the lattice data ! We thus conclude that the SU(6) assumption does not provide us with a reasonable mechanism that would allow for an enhancement of the chiral extrapolation curve around the physical pion mass. Similar conclusions can be drawn from the recent calculation by Detmold et al. \cite{Detmold2}, which also utilized SU(6) assumptions to reduce the number of unknown couplings. Interestingly, we note that the resulting value for $C(1\rm{GeV})$ is still rather small, but {\em has changed sign} when compared to the free fit results. In the following we will pursue the hypothesis that it is the sign and magnitude of this coupling (and the resulting interplay with the non-analytic chiral results given in Eq.(\ref{gasse})) that are responsible for the enhancement of $g_A(m_\pi)$ around physical pion mass.  

\subsection{Fit Constrained by $\pi N\rightarrow\pi\pi N$}
When varying the three poorly known parameters $g^0_A$, 
$C(\lambda),\,g_1$ of the SSE extrapolation formula Eq.(\ref{gasse}) 
one realizes that it is the short-distance physics encoded in $C(\lambda)$ which is crucial for a possible enhancement in $g_A(m_\pi)$ at small pion masses. Empirically one finds that with negative values around for $C(\lambda=1\rm{GeV})\approx -4$ GeV$^{-2}$ it would be possible to counterbalance the trend of $g_A(m_\pi)$ towards values below 1, which is driven by the chiral logarithm (cf. fit Ia in Fig.\ref{su6fit}). The question arises whether such a value for $C(\lambda=1\rm{GeV})$ at such a regularization scale is consistent with known physics.  

At first glance one would expect that only lattice simulations can provide more information on $C(\lambda)$---respectively the couplings $B_9,\;B_{20}$---to further constrain the fits. However, when analyzing the chiral tensors in Eq.(\ref{counterterms}), one notices that the physically observable coupling $B_9$ also contributes in inelastic pion-nucleon scattering processes like $\pi N\rightarrow \pi\pi N$ \cite{Triumf}. In ref.\cite{nadia1} this process was analyzed to ${\cal O}(p^3)$ in HBChPT and values for the couplings of interest were obtained\footnote{A second analysis of $\pi N\rightarrow\pi\pi N$ scattering \cite{Regina} unfortunately does not specify the value for the analogue of the coupling $B_9$ they utilized.}. Here, however, we follow the substantially revised analysis of ref.\cite{nadiathesis}. Translating these findings into our conventions we obtain
\begin{eqnarray}
B_9^{r}(\lambda=m_\pi^{phys})=\left(-1.4\pm1.2\right){\rm GeV}^{-2}\;,\quad B_{20}^{r}(\lambda=m_\pi^{phys})\equiv 0\;,\label{couplings}
\end{eqnarray}
with the error bar arising from three different fitting procedures described in \cite{nadiathesis}. We note that refs.\cite{nadia1,nadiathesis} follow the convention to set all equation-of-motion dependent counterterms like $B_{20}$ identically equal to zero {\em at the scale of their analysis}, which corresponds to $\lambda=0.138$ GeV. While we can directly utilize this information in the heavy baryon analysis (c.f. the discussion in section \ref{5}) there is no unique procedure how to import this information into the framework with explicit $\Delta$(1230) degrees of freedom. The complication arises from the fact that $B_9^{r}(\lambda),\,B_{20}^{r}(\lambda)$ have different beta-functions in HBChPT and SSE. One can therefore produce different chiral extrapolation curves depending on the scale where the matching between the two theories takes place. In the following we will perform the matching at the intermediate scale $\lambda=2\Delta_0=0.54$ GeV which suggests itself as a ``natural'' scale in SSE by looking at Eqs.(\ref{decoupling},\ref{gachiral}). We have carefully examined that our results are not sensitive (within the uncertainties of the $\pi N \to \pi \pi N$ analysis) to changes of the scale $\lambda$ between 0.4 and 0.8 GeV. At the end of this section we demonstrate that this matching choice is consistent with available information on the chiral limit behaviour of $g_A$, which is the only constraint one can put on different matching prescriptions.
First, we evaluate the couplings of Eq.(\ref{couplings}) at the scale $\lambda=2\Delta_0$ via the relation
\begin{eqnarray}
B_i^{r}(2\Delta_0)=B_i^{r}(m_\pi^{phys})-\frac{\beta_i^{HB}}{(4\pi f_\pi)^2}\ln\frac{2\Delta_0}{m_\pi^{phys}}\;,\quad i=9,20
\end{eqnarray}
leading to the values
\begin{eqnarray}
B_9^{r}(\lambda=2\Delta_0)|_{HB}=\left(-1.8\pm 1.2\right){\rm GeV}^{-2}\;,\quad B_{20}^{r}(\lambda=2\Delta_0)|_{HB}=0.9\, {\rm GeV}^{-2}\;.\label{HBcouplings}
\end{eqnarray}
The leading-one-loop HBChPT beta-functions required can be found in ref.\cite{Ecker}. Note that we have utilized physical couplings in the heavy baryon beta-functions, inducing an uncertainty which is negligible compared to the error bar in $B_9$. At the scale $\lambda=2\Delta_0$ we then perform the matching
\begin{eqnarray}
B_9^{r}(\lambda=2\Delta_0)|_{SSE}\equiv B_9^{r}(\lambda=2\Delta_0)|_{HB}\,,\quad
B_{20}^{r}(\lambda=2\Delta_0)|_{SSE}\equiv B_{20}^{r}(\lambda=2\Delta_0)|_{HB}\; . \label{matching}
\end{eqnarray}
Next we run the SSE couplings up to the scale $\lambda=1$ GeV to facilitate an easier comparison with the previous fits (c.f. table \ref{tab2}):
\begin{eqnarray}
B_i^{r}(1{\rm{GeV}})=B_i^{r}(2\Delta_0)-\frac{\beta_i^{SSE}}{(4\pi f_\pi)^2}\ln\frac{1{\rm{GeV}}}{2\Delta_0}\;,\quad i=9,20\; .
\end{eqnarray}
Utilizing the leading-one-loop SSE beta-functions of ref.\cite{BFHM} with the couplings shown in fit III of table \ref{tab2} one finally obtains
\begin{eqnarray}
B_9^{r}(\lambda=1{\rm{GeV}})|_{SSE}=\left(1.2\pm 1.2\right){\rm GeV}^{-2}\;,\quad B_{20}^{r}(\lambda=1{\rm{GeV}})|_{SSE}=1.9\, {\rm GeV}^{-2}\;.\label{GeVcouplings}
\end{eqnarray}
The error bar given for $B_9(1\rm{GeV})$ originates from the determination of ref.\cite{nadiathesis} shown in Eq.(\ref{couplings}). No errors arising from the couplings in the HBChPT or SSE beta functions are shown, as they are smaller than the uncertainty in $B_9$ resulting from the $\pi N\rightarrow\pi\pi N$ analyses. 

\begin{figure}[t]
  \begin{center}
    \includegraphics*[width=0.7\textwidth]{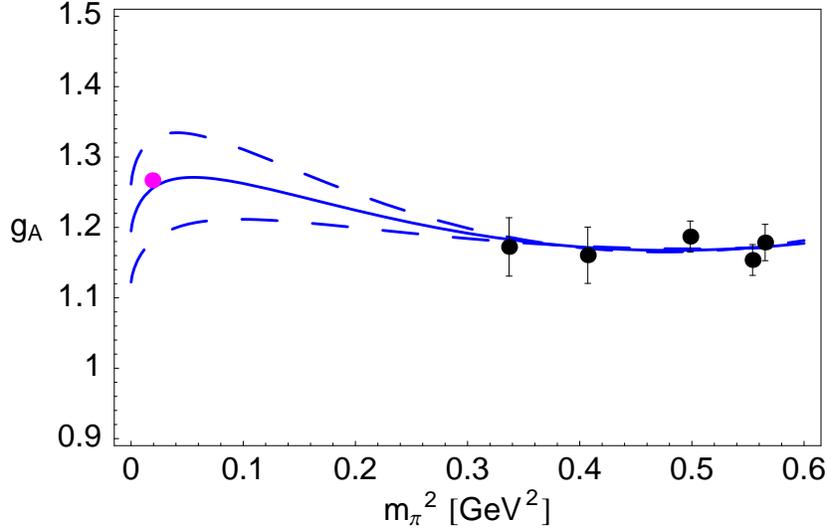}
    \caption{Fit III: Incorporating known information from $\pi N\rightarrow\pi\pi N$ differential cross sections into the leading-one-loop SSE analysis. The solid line shows the central value, whereas the dashed curves denote the upper and lower error bar from Eq.(\ref{GeVcouplings}). The  matching of the coupling constants was performed at $\lambda=0.54$ GeV as discussed in the text.}\label{cfit}
  \end{center}
\end{figure}

\begin{figure}[t]
  \begin{center}
    \includegraphics*[width=0.7\textwidth]{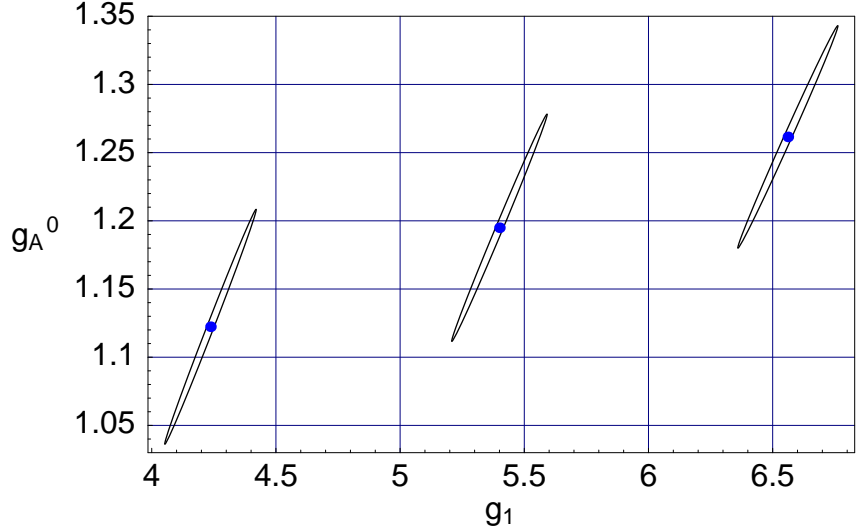}
    \caption{Confidence ellipses for the constrained leading-one-loop SSE fit III shown in Fig.\ref{cfit}. The 2 parameters shown denote the nucleon coupling constant $g_A^0$ in the chiral limit versus the axial $\Delta \Delta$ coupling $g_1$. The three ellipses shown correspond to central, upper and lower values for the short distance couplings determined in Eq.(\ref{GeVcouplings}) at a scale of $\lambda=1$ GeV.}\label{cfitconf}
  \end{center}
\end{figure}

The resulting chiral extrapolation curve for $g_A(m_\pi)$ utilizing Eqs.(\ref{gasse},\ref{GeVcouplings}) and fitting $g_A^0,\,g_1$ to the five lattice data points is shown in Fig.\ref{cfit}. The resulting parameters are shown as fit III in table \ref{tab2}. The solid line in Fig.\ref{cfit} refers to the central value for $B_9$ in Eq.(\ref{GeVcouplings}), whereas the two dashed curves indicate the influence of the (large) uncertainty about the precise value of $B_9$ given in Eq.(\ref{GeVcouplings}). Nevertheless, while the $\chi^2$/d.o.f. values are comparable with fits I and II shown in Fig.\ref{su6fit}, only fit III is consistent with the physical point $g_A^{phys}$ as shown in Fig.\ref{cfit}. We note that it is indeed the large negative values for the parameter 
\begin{eqnarray}
C(1{\rm{GeV}})=B_9^{r}(1{\rm{GeV}})-2\,g_A^0\,B_{20}^{r}(1{\rm{GeV}})=(-3.4\pm 1.2)\;\rm{GeV}^{-2}\label{C1GeV}
\end{eqnarray}
that can produce an enhancement near the physical point. Given the large uncertainty in $C(1\rm{GeV})$ (resulting from $B_9$) we believe that at present any normalization issues (e.g. resulting from finite volume effects) can be accounted for by our extrapolation function $g_A(m_\pi)$ of Eq.(\ref{gasse}). Analyzing the ellipsoidal joint 95\% confidence region for the fit parameters $(g_A^0,g_1)$ associated with the central value of $C(1{\rm GeV})$ in fit III (c.f. Fig.\ref{cfitconf}), the band of allowed fit curves completely covers the uncertainty region resulting from $B_9$ (dashed curves) and shows that at this confidence level no fit solutions exist for which $g_A(m_\pi)$ becomes smaller than unity, in agreement with the Adler-Weisberger sum rule Eq.(\ref{AW}).

We therefore conclude that the strong downward-bending of $g_A(m_\pi)$ driven by the Goldstone boson cloud around spin 1/2 and spin 3/2 baryon states is counterbalanced by large short-distance corrections parameterized via $C(\lambda)$. The combination of these two effects leads to a rather flat chiral extrapolation function for $g_A(m_\pi)$, which---according to crucial input from $\pi N\rightarrow\pi\pi N$ scattering analyses---can display an enhancement near the physical point to reconcile the lattice data with our knowledge from neutron beta decay. The analysis presented here suggests that in the (lattice) observable $g_A(m_\pi)$ no strong curvature is to be expected between the physical point and the location of the lattice data. The large LNA terms contained in $g_A(m_\pi)$ presumably only become visible for quark masses rather close to the chiral limit.

Finally, we comment on the validity of our matching procedure resulting in the couplings of Eq.(\ref{GeVcouplings}). Once the couplings are specified at one particular scale the resulting SSE extrapolation curve shown in Fig.\ref{cfit} does not depend anymore on $\lambda$, but it does depend on the matching point Eq.(\ref{matching}). Nevertheless, for each baryon observable there exists one unique curve through which the chiral limit value is approached when the quark masses go to zero. This is the reason why it is possible to utilize information about coupling constants in one version of ChEFT and transcribe it to another version of that theory. From Eq.(\ref{gachiral}) we already know that the first two terms in the chiral expansion of $g_A(m_\pi)$ are identical in HBChPT and SSE, providing an important consistency check. However, the terms $\sim m_\pi^2$ can be different between the two ChEFTs---even when each one is evaluated at leading-one-loop order as done here---because the two schemes can build up the contributions to a particular term in the chiral expansion more or less effectively in their respective perturbative series. Utilizing Eq.(\ref{HBcouplings})
one finds
\begin{eqnarray}
g_A^\chi|_{HB}&=&1.2+4\,\gamma^{HB}\,m_\pi^2\ln
                          \frac{m_\pi}{2\Delta_0}+m_\pi^2
                          (-17.8\pm 4.8)
                          {\rm{GeV}^{-2}}+{\cal O}(m_\pi^3)\;\label{hbchir}
\end{eqnarray}
for the chiral limit behaviour of the leading-one-loop HBChPT result of Eq.(\ref{gAA}). On the other hand, utilizing the parameters of fit III (c.f. table \ref{tab2}) one obtains
\begin{eqnarray}
g_A^\chi|_{SSE}&=&1.2+4\,\gamma^{HB}\,m_\pi^2\ln
                          \frac{m_\pi}{2\Delta_0}+m_\pi^2
                          (-10.4\pm 4.8)
                          {\rm{GeV}^{-2}}+{\cal O}(m_\pi^3)\;
\end{eqnarray}
for the chiral limit behaviour of the leading-one-loop SSE result of Eq.(\ref{gasse}). The $m_\pi^2$ terms agree within the error bars and the shifts of the central values could reasonably be accounted for by higher order corrections to the heavy baryon result of Eq.(\ref{hbchir}). We conclude that the matching procedure used in this work is consistent with all presently known information about the chiral limit of $g_A(m_\pi)$. In the next section we will test the stability of fit III by including the physical point as an additional constraint for the fit and then conclude our analysis.

\section{Discussion}\label{5}

\begin{figure}[t]
  \begin{center}
    \includegraphics*[width=0.7\textwidth]{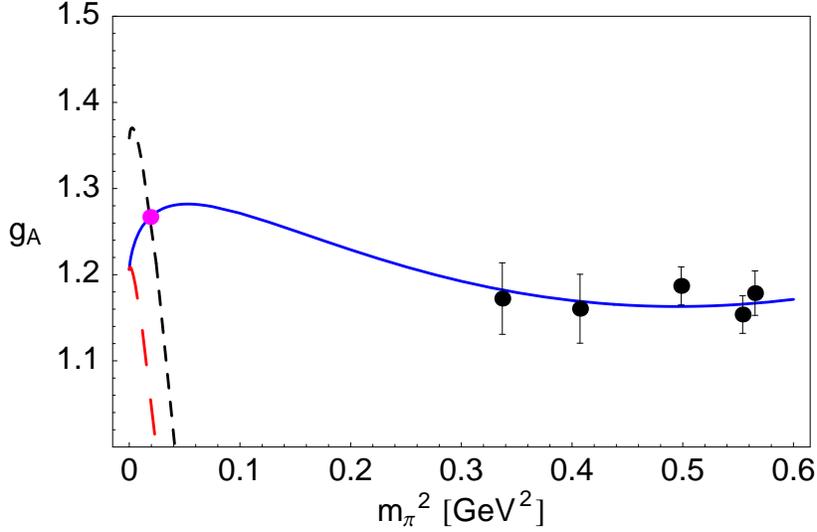}
    \caption{Fit IV: Incorporating the physical point and the lattice data into the leading-one-loop SSE extrapolation formula results in a well behaved chiral extrapolation function (solid line, fit IVb) from the chiral limit over the physical point out to the region of the lattice data. The short-dashed line shows the corresponding leading-one-loop HBChPT result (fit IVa). The long-dashed line (fit IVc) indicates the HBChPT result assuming the same chiral limit value $g_A^0$ as in the SSE analysis of fit IVb.}\label{bestfit}
  \end{center}
\end{figure}

In the previous section we have found that the leading-one-loop SSE result Eq.(\ref{gasse}) together with input from scattering analyses can provide a meaningful chiral extrapolation of the present lattice data which is consistent with the experimentally known value for $g_A$. In order to test the stability of our result we now study the changes to our curves when we include the information at the physical point as an additional constraint together with the lattice data. Reassuringly, one finds only small changes in the SSE case. Numerical values for the set of couplings labeled fit IVb are given in table \ref{tab2}, entirely consistent with the couplings of fit III which does not know about the location of the physical point. The resulting extrapolation function is shown in Fig.\ref{bestfit}, providing a sensible connection between the chiral limit value, the physical point and the QCDSF lattice data. We note that a large negative value of $C(1\rm{GeV})$ and $g_1>g_1^{SU(6)}$ are common features of fits III and IVb, indicating that our reasoning provided in the previous section was on the right track.

\begin{figure}[t]
  \begin{center}
    \includegraphics*[width=0.7\textwidth]{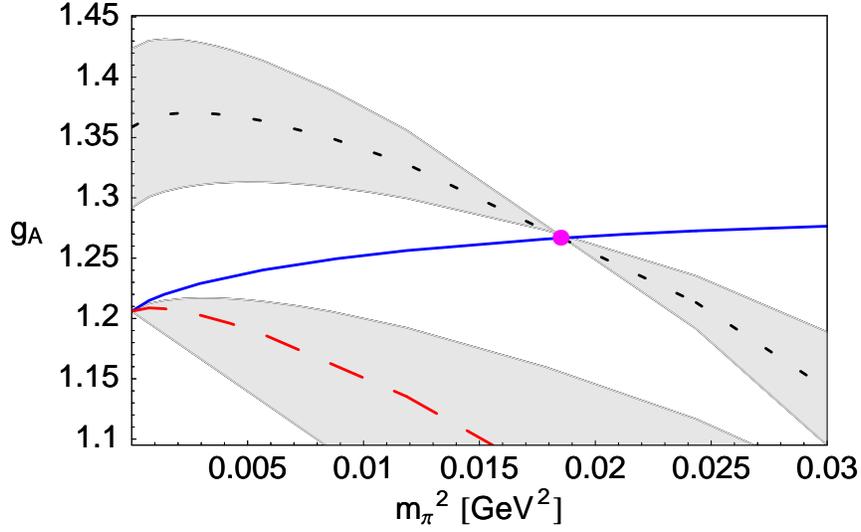}
    \caption{We have magnified the region around the chiral limit up to the physical point. The bands indicated for the two heavy baryon curves arise from the uncertainty of the $\pi N\rightarrow\pi\pi N$ analysis of Eq.(\ref{couplings}).}\label{bestfit2}
  \end{center}
\end{figure}

When taking the physical point as additional input, we can now also examine the validity of the leading-one-loop HBChPT extrapolation formula of Eq.(\ref{gAA}). In fit IVa we have taken $C(1\rm{GeV})=-5.3$ GeV$^{-2}$, which corresponds to the (central) value suggested in ref.\cite{nadiathesis}. The resulting (short-dashed) curve shown in Fig.\ref{bestfit} completely misses the lattice data and leads to a substantially different chiral limit value $g_A^0$, even if the rather large uncertainty in $C(1\rm{GeV})$ (c.f. table \ref{tab2})---indicated via the bands in Fig.\ref{bestfit2}---is taken into account. According to our analysis this finding forces us to the conclusion that the leading-one-loop HBChPT curve as given in Eq.(\ref{gAA}) is not sufficient to even perform the chiral extrapolation from the chiral limit to the physical point! With the help of the decoupling constraints Eq.(\ref{decoupling}) we have made sure that the chiral limit value of $g_A(m_\pi)$ is the same in both approaches, as there exists only one definite axial charge of the nucleon in the chiral limit. In order to further test the applicability of the HBChPT curve we have also performed fit IVc, demanding that $g_A^0$ takes the same value as in the leading-one-loop SSE analysis. The resulting (long-dashed) extrapolation curve is shown in Fig.\ref{bestfit}. It only follows the SSE curve for extremely small values of $m_\pi$, misses the physical point and breaks down before it comes anywhere close to the existing lattice data. We therefore consider the use of Eq.(\ref{gAA}) as not appropriate for a stable chiral extrapolation outside the region of the extreme chiral limit. 

We note that it is the very leading-one-loop HBChPT formula Eq.(\ref{gAA}) that has been utilized recently in refs.\cite{silas,efgeni} to study the behaviour of the deuteron in the chiral limit. In view of the findings presented here the question arises whether conclusions drawn in refs.\cite{silas,efgeni} regarding the bound state nature of the deuteron are unaltered, when the chiral extrapolation of $g_A(m_\pi)$ from the physical point down to the chiral limit follows the curve from fit IVb, rather than the one from IVa as shown in Fig.\ref{bestfit}.  

\section{Summary and Outlook}\label{6}

To conclude, we summarize the main results of our work:
\begin{itemize}
\item[-] The leading-one-loop HBChPT formula for $g_A(m_\pi)$ Eq.(\ref{gAA}) depends on two free parameters $g_A^0,\,C(\lambda)$. Crucial information on sign and size of $C(\lambda)$ can be obtained from analyses of the $\pi N\rightarrow \pi\pi N$ reaction. $C(\lambda)$ is large and negative for typical values of the regularization scale.
\item[-] The leading-one-loop SSE formula for $g_A(m_\pi)$ Eq.(\ref{gasse}) can be matched to the HBChPT formula in the chiral limit. It has sufficient quark mass dependent structures to provide a consistent chiral extrapolation from the chiral limit across the physical point up to the lattice data, assuming that the large quark mass quenched lattice data are consistent with fully dynamical simulations and that finite volume effects do not provide large corrections. 
\item[-] Present lattice data are not precise enough in order to determine all couplings of the SSE formula from the data alone. Additional information from $\pi N$ dynamics has to be invoked. SU(6) arguments are not sufficient to generate the required form of $g_A(m_\pi)$. 
\item[-] The chiral extrapolation curve for $g_A(m_\pi)$ is rather flat. It implies a co-operation of terms in the quark mass expansion such that pion cloud effects are basically ``turned off'' for pion masses exceeding $0.6$ GeV. A small enhancement is seen around the physical pion mass. This enhancement results from an interplay between the Goldstone boson dynamics of the nucleon and short distance contributions which can be constrained from $\pi N\rightarrow \pi\pi N$ scattering data. The chiral limit value of the axial coupling of the nucleon is found to be 

\begin{eqnarray}
g_A^0\approx 1.2\pm 0.1\,.
\end{eqnarray}
 
The parameters of fit IVb give the best available description of our present knowledge on $g_A(m_\pi)$. 
\item[-] The leading one-loop HBChPT formula (\ref{gAA}) for $g_A(m_\pi)$ is not useful for chiral extrapolation. It fails to connect the chiral limit with the physical value of the pion mass. According to our analysis the chiral logarithm as the leading non-analytic quark mass term in the HBChPT formula is only visible in extrapolations performed {\em below} the physical point.
\end{itemize}
Finally, we note that the role of finite size effects in the lattice data for $g_A$  has to be analyzed in a systematic framework to better understand the systematics of the lattice error bars. More data from simulations for $g_A(m_\pi)$ performed at different lattice volumes are called for to guide these studies. We are confident that ChEFT is an important tool to resolve this issue in the future.

\begin{center}
{\bf Acknowledgments}
\end{center}

The authors acknowledge helpful discussions with F. Arleo, M. G{\" o}ckeler,
H. Grie{\ss}hammer, N. Kaiser, E. Kolomeitsev, U.-G. Mei{\ss}ner, W. Melnitchouk, A. Sch{\" a}fer, G. Schierholz and A.W. Thomas. We are grateful to the QCDSF collaboration for providing 
us with their simulation data prior to publication.
 TRH would like to acknowledge the hospitality of ECT* in Trento.
This work was supported in part by BMBF and DFG.
\newpage

\appendix

\section{Amplitudes}\label{A}

Here we present the results for the 8 leading-one-loop amplitudes, {\it i.e.} ${\cal O}(\epsilon^3)$ diagrams shown in Fig.\ref{Feyndiags} which can contribute\footnote{To simplify the calculation we have made use of the gauge-condition $\epsilon_A\cdot q=0$.} to the axial coupling constant $g_A$  of a nucleon. The Lagrangians needed for this calculation are discussed in the main text. We work in the Breit-frame and choose the velocity vector $v^\mu=(1,0,0,0)$. With $S^\mu$ denoting the Pauli-Lubanski spin-vector and $\epsilon_A^\mu$ the polarization 4-vector of an external axial-vector source, one finds
\begin{eqnarray}
Amp_{1}&=&i\;\eta^{\dag}\tau^i\eta\;\bar{u}(r_2)S\cdot\epsilon_Au(r_1)\left(g^0_AZ_N^{SSE}+4m_\pi^2B_9+\Delta_0^2 B_{31}\right)\\
Amp_{2}&=&Amp_{3}=0+{\cal O}(\epsilon^4)\nonumber\\
Amp_{4}&=&i\frac{(g^0_{A})^3}{f_\pi^2}\;\eta^{\dag}\tau^{i}\eta\;\frac{\partial}{\partial r_0}J_2(r_0)\mbox{\Large $|$}_{r_0=0}\;\bar{u}(r_2)S\cdot\epsilon_Au(r_1)\left(\frac{1}{2}+S^2\right)\\
Amp_{5}&=&-i\frac{g^0_{A}}{(4\pi f_\pi)^{2}}\;\eta^{\dag}\frac{\tau^{i}}{2}\eta\;\bar{u}(r_2)S\cdot\epsilon_Au(r_1)\;4m_\pi^2\left(16\pi^2L+\ln{\frac{m_\pi}{\lambda}}\right)\\
Amp_{6}&=&Amp_{7}=-i\;\frac{g^0_Ac_A^2}{\Delta f_\pi^2}\;\frac{8}{3}\;\eta^{\dag}\tau^i\eta\;\frac{d-2}{d-1}\;\bar{u}(r_2)S\cdot\epsilon_Au(r_1)[J_2(0)-J_2(-\Delta_0)]\\
Amp_{8}&=&-i\frac{10g_1c_A^2}{9f_\pi^2}\eta^{\dag}\tau^i\eta\left[d-3+\frac{(d-3)^2}{(d-1)^2}\right]\bar{u}(r_2)S\cdot\epsilon_Au(r_1)\frac{\partial}{\partial r_0}J_2(r_0-\Delta_0)\mbox{\Large $|$}_{r_0=0}\label{amplitudes}
\end{eqnarray}
Explicit expressions for the function $J_2$ are given in \cite{Compton}. We
evaluate the amplitudes in $d$ dimensions with induced regularization scale
$\lambda$. Any ultraviolet divergences appearing in the limit $d\rightarrow 4$
are subsumed in 
\begin{eqnarray}
L=\frac{\lambda^{d-4}}{16\pi^2}\left[\frac{1}{d-4}+\frac{1}{2}\left(\gamma_E-1-\ln 4\pi\right)\right],
\end{eqnarray}
where $\gamma_E$ denotes the Euler-Mascheroni constant.

Finally, we note that with the decoupling choice made in Eq.(\ref{decoupling}) the nucleon Z-factor calculated to leading-one-loop in SSE reads
\begin{eqnarray}
Z_N^{SSE}&=&1-\frac{1}{(4\pi f_\pi)^2}\left\{\left[\frac{3}{2}\,(g_A^0)^2+4\,c_A^2\right]m_\pi^2\right.\nonumber\\
         & &+8\,m_\pi^2\left[16\pi^2 f_\pi^2\,B_{20}^{r}(\lambda)+\left(\frac{9}{16}\,(g_A^0)^2+c_A^2\right)\ln\frac{m_\pi}{\lambda}\right]\nonumber\\
         & &\left.-16c_A^2\Delta_0^2\left[\ln\left(\frac{m_\pi}{2\Delta_0}\right)+\sqrt{1-\frac{m_\pi^2}{\Delta_0^2}}\ln R\right]\right\}
+{\cal O}(\epsilon^4)\;,
\end{eqnarray}
where $R$ has been defined in Eq.(\ref{R}).
\newpage


\end{document}